\documentclass[aps,prl,reprint,superscriptaddress]{revtex4-1}
\usepackage{blindtext}
\usepackage{graphicx} 
\usepackage{xcolor}
\usepackage{physics}
\usepackage{amsmath,amssymb,amsfonts}
\usepackage{braket}
\usepackage[utf8]{inputenc}
\usepackage[T1]{fontenc}
\usepackage[colorlinks=true,linkcolor=blue,citecolor=blue,urlcolor=blue]{hyperref}
\usepackage{amsthm}

\newtheorem{lemma}{Lemma}

\begin{document}

\title{Big cats: entanglement in 120 qubits and beyond}

\makeatother
\author{Ali Javadi-Abhari}
\email{ali.javadi@ibm.com}
\affiliation{IBM Quantum, T. J. Watson Research Center, Yorktown Heights, NY}

\author{Simon Martiel}
\affiliation{IBM Quantum, IBM France Lab, Saclay, France}

\author{Alireza Seif}
\affiliation{IBM Quantum, T. J. Watson Research Center, Yorktown Heights, NY}

\author{Maika Takita}
\affiliation{IBM Quantum, T. J. Watson Research Center, Yorktown
Heights, NY}

\author{Ken X. Wei}
\affiliation{IBM Quantum, T. J. Watson Research Center, Yorktown Heights, NY}

\begin{abstract}

Entanglement is the quintessential quantum phenomenon and a key enabler of quantum algorithms. The ability to faithfully entangle many distinct particles is often used as a benchmark for the quality of hardware and control in a quantum computer. Greenberger–Horne–Zeilinger (GHZ) states, also known as Schr\"odinger cat states, are useful for this task. They are easy to verify, but difficult to prepare due to their high sensitivity to noise. In this Letter we report on the largest GHZ state prepared to date consisting of 120 superconducting qubits. We do this via a combination of optimized compilation, low-overhead error detection and temporary uncomputation. We use an automated compiler to maximize error-detection in state preparation circuits subject to arbitrary qubit connectivity constraints and variations in error rates. We measure a GHZ fidelity of $0.56(3)$ with a post-selection rate of $28\%$. We certify the fidelity of our GHZ states using multiple methods and show that they are all equivalent, albeit with different practical considerations.

\end{abstract}

\maketitle

Entanglement is a defining feature that separates quantum physics from classical physics, allowing correlations to exist among multiple particles that cannot be described by their local properties. GHZ states~\cite{greenberger1989going}, related to the famous Schr\"odinger's cat thought experiment, are an important class of quantum states exhibiting maximal correlation among all particles. These states are incredibly useful, finding application in areas such as quantum secret sharing~\cite{hillery1999quantum}, quantum metrology~\cite{Giovannetti2004}, gate teleportation~\cite{gottesman1999demonstrating} and error correction~\cite{shor1995scheme}.

Beside their practical utility, GHZ states have historically been used as a benchmark in various quantum platforms such as ions~\cite{sackett2000experimental,leibfried2005creation,monz201114,q2b}, superconductors~\cite{song201710,wei2020verifying,mooney2021generation,bao2024creating,liao2025achieving}, neutral atoms~\cite{omran2019generation,song2019generation} and photons~\cite{wang201818}. This arises from the fact that these states are extremely sensitive to imperfections in the experiment~\cite{gottesman2019maximally}---indeed they can be used to achieve quantum sensing at the Heisenberg limit~\cite{bollinger1996optimal}. Successful preparation of large GHZ states can therefore serve as a demonstration of good control and suppression of noise, which is essential for large-scale quantum computation. Here we establish the largest GHZ states reported to date, consisting of 120 qubits (protected by an additional 8 qubits) in a fixed-frequency, tunable-coupler superconducting processor, extending the previous record~\cite{liao2025achieving} by 45 qubits. 

Our experiment is enabled by a tailored implementation of the state preparation that manages to suppress different sources of errors. We use the structure present in GHZ states to generate them from the ground up within a novel compiler framework that adaptively compiles the circuit to utilize the best regions of the device, incorporate syndrome measurements dynamically within arbitrary connectivity constraints to ensure as many errors are flagged as possible, and increase the coherence by using temporary uncomputation.

Verifying the fidelity of large quantum states is generally challenging due to the extensive number of measurements required. This difficulty is further exacerbated by the use of error detection and post-selection techniques, which, while improving the quality of retained output samples, effectively slow down the quantum computer. Additionally, readout errors introduce ambiguity in distinguishing between errors that genuinely affect entanglement and those that only impact measurement outcomes without degrading the prepared state's quality.
Traditionally, GHZ states have been verified using Multiple Quantum Coherences (MQC)~\cite{wei2020verifying} and parity oscillation tests~\cite{guhne2007toolbox}, both of which require $2N$ distinct experiments, each with a sufficient number of shots to accurately estimate fidelity.
In this work, alongside parity oscillation tests, we employ Direct Fidelity Estimation (DFE)~\cite{flammia2011direct,da2011practical}, a method applicable to any stabilizer state (including GHZ as a special case). DFE requires only a constant number of measurements, offering a more scalable approach. These methods are affected differently by readout errors and therefore need tailored mitigation strategies. We establish a concrete connection between the measurements used in both approaches and demonstrate excellent agreement in experimental results, further validating our readout error mitigation strategy.

\begin{figure*}[t]
\centering
\includegraphics[width=\textwidth]{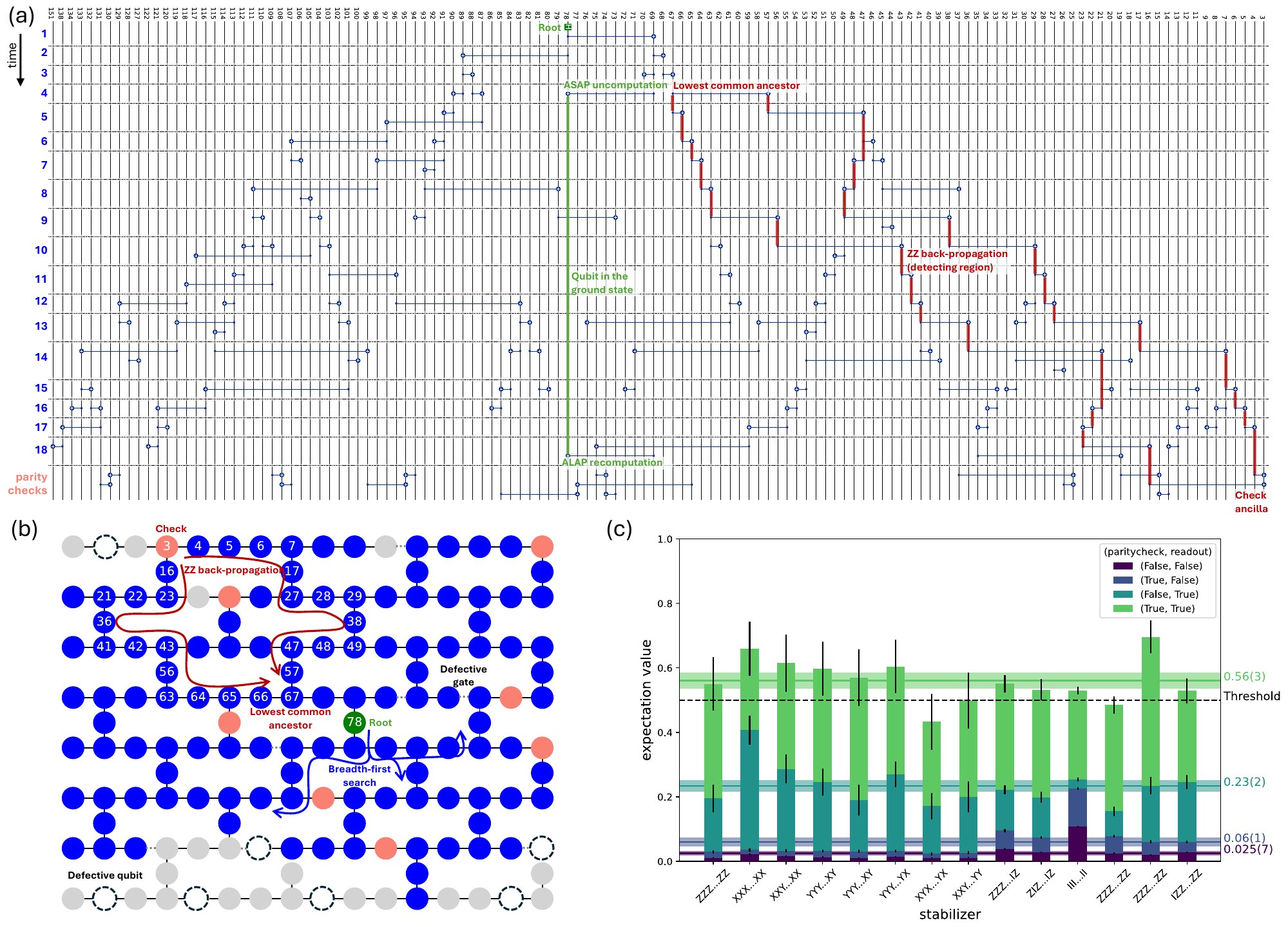} 
\caption{Preparation of a 120-qubit GHZ state on the $ibm\_aachen$ processor. (a) The circuit has a CNOT depth of 18, with a final layer of parity checks consisting of 8 ancilla qubits. The spacetime detecting region of one of the ancilla qubits is highlighted in red, consisting of all edges that lie in the subtree spanning from the checked qubits to their lowest common ancestor. (b) The same region is highlighted spatially on the layout of qubits, where the GHZ state begins from the root qubit highlighted in green and spreads outwards in blue in a breadth-first-search manner. Missing nodes and edges signify dropouts due to high gate or measurement error. The position of ancillas are highlighted in pink. The root qubit is uncomputed early in the circuit, and only recomputed in the final layer, resulting in a long timespan highlighted in green where the qubit is relaxed and in the ground state. (c) To certify the state fidelity, we use DFE and measure seven diagonal and seven non-diagonal stabilizers, with their average fidelities and error bars shown for various cases of parity check or readout error mitigation. 
}\label{fig:120}
\end{figure*}

\section{Adaptive compilation of GHZ states}

We seek to create a large entangled resource state on a quantum computer using a circuit whose noise is suppressed. We use techniques from graph theory, stabilizer groups, and circuit uncomputation to achieve this goal. 

An $n$-qubit GHZ state is given by 
\begin{equation}
    \ket{G_n} = \frac{1}{\sqrt{2}}(\ket{0}^{\otimes n} + \ket{1}^{\otimes n}).
\end{equation}

This state is invariant under qubit permutation, and moreover it admits low-weight stabilizers in the form of ${Z_i Z_j}$ for arbitrary pairs of qubits $(i, j)$. These are the key properties that we exploit in our compilation framework, which is illustrated in Figure~\ref{fig:120}.

Our circuits incorporate error detection, which trades off speed for accuracy by discarding executions in which we can detect the presence of errors. This is a single-shot form of error mitigation, in the sense that those runs that are accepted do in fact prepare a high-fidelity quantum state that can be consumed to perform further computation.

A key practical challenge in experimental quantum computing is the non-uniformity of qubits and gates, where a few low-performing ``tails'' can render an otherwise high-performing regular lattice of qubits into an irregular lattice. These can be attributed to device yield, presence of transient two-level systems (TLS), or other physical effects. We broadly refer to these as defective qubits and gates (the definition is arbitrary and subject to a performance cutoff threshold). This means that traditional approaches to circuit compilation, which first build a circuit then map it to hardware~\cite{javadi2024quantum,nation2023suppressing}, can fail for large-scale circuits. The problem is further exacerbated when ancilla qubits must be allocated physically close to data qubits for error detection. We therefore choose to adaptively grow the circuit on the hardware in order to meet various criteria directly.

\paragraph{Adaptive tree construction ---} Consider a graph $ G(V,E) $ that represents the connectivity of qubits in the hardware, whose vertices and edges are the qubits and native two-qubit gates, respectively. By the definition of the GHZ state, the order in which the qubits are entangled is unimportant. A short-depth GHZ state can be prepared on G by initiating the entangling sequence from a vertex of minimal eccentricity, defined as
\[
e(u) = \max \{\, d(u, v) \mid v \in V \,\},
\]
where \( d(u, v) \) denotes the shortest-path distance between \( u \) and \( v \). Placing such a root \(v_0\) in a superposition state and performing a breadth-first-search (BFS)~\cite{bundy1984breadth} with $CNOT$ gates along the edges of $G$, until the desired number of vertices are visited, results in a GHZ state with circuit depth 
\begin{align}
   D_{\mathrm{GHZ}} \propto e(v_0) = \min_{u \in V} e(u) \leq \mathrm{radius}(G). 
\end{align}

A circuit constructed as such, while short-depth, is unlikely to admit good checks, since we have not accounted for the availability of ancilla yet. Next we describe a procedure for scoring available parity checks, and a randomized algorithm that perturbs the circuit to optimize this score.

\paragraph{Detecting regions via $ZZ$ back-propagation ---}

The structure of a GHZ state naturally allows for low-overhead parity checks that can flag errors in the circuit. By measuring the $ZZ$ parity of any two qubits, we can detect violations from the expectation of even parity and discard the prepared state as invalid. This was used in Refs.~\cite{mooney2021generation,liao2025achieving} to successfully improve the fidelity of GHZ states.

However, not all parity checks are the same. We use tools similar to those developed in Ref.~\cite{martiel2025low} to substantially improve the checks' performance by maximizing their detecting regions. However, unlike previous work, we do not fix the location of checks and instead allocate them dynamically during circuit preparation to make them both effective and compatible with the hardware graph $G$.


We use the following Lemma (see proof in the Appendix and Figure~\ref{fig:120} for an illustration).

\begin{lemma}[Detecting region of a GHZ parity check]
Let $C$ be a GHZ-preparation circuit whose gates and idle timesteps form a rooted tree $T=(V,E)$.
Define the edge set
\begin{align}
S_{i,j}
\;:=\;
\mathrm{path}(i,\mathrm{lca}(i,j))
\;\cup\;
\mathrm{path}(j,\mathrm{lca}(i,j))
\;\subseteq\; E,
\end{align}

where \(\mathrm{lca}(i,j)\) is the lowest common ancestor of leaves \(i,j\in V\). 
A single-qubit $X$ or $Y$ error at a spacetime location $(q, t)$ in $C$ is detected by the
measurement of \(Z_i Z_j\) if and only if its corresponding edge in $T$ lies in $S_{i,j}$. 
\end{lemma}

This lemma states that a check is sensitive to errors that lie on the paths connecting the checked qubits to their lowest common ancestor, which is the unique sequence of edges connecting each leaf upward until their trajectories first intersect (see Fig.~\ref{fig:120}). We define the ``coverage'' of multiple checks as the ratio of spacetime locations detected by at least one of them, to the total number of spacetime locations in the circuit where an error can occur.

Note that the coverage objective naturally balances other considerations that one might have about the circuit. For example, naively one would expect that more checks are better; however this may increase circuit depth and introduce more fault locations, due to check ancilla blocking certain paths on chip. Therefore, rather than separately optimizing circuit depth or number of checks, we optimize the coverage fraction of checks over the circuit. We find that this metric explains the performance of different checks well. For example, Liao et al.~\cite{liao2025achieving} considered a type of check that was attached to two sequential qubits, and concluded that this check does not work well due to needing an extra swap in its implementation. However, from our framework it is immediately obvious why the check does not perform well: those two qubits have a trivial detecting region, as their lowest common ancestor is included in them. Instead, we expect checks to perform best when two long branches of the GHZ tree converge onto the same ancilla.

It remains to grow the circuit in a way that optimizes coverage as defined above. To do this we randomly block certain nodes during the BFS from the root outwards. This forces the GHZ support to go around certain nodes in the hardware graph G, and creates more ancilla space. After the GHZ is prepared, any two nodes in its support that neighbor a common ancilla are used for a parity check. We compute their coverage, and repeat the randomized trial. The best-covered circuit is chosen at the end.


\paragraph{Temporary uncomputation ---}
The parity checks introduced will only detect $X$ or $Y$ errors. While this covers cases such as gate errors or spontaneous qubit relaxations, it leaves out Z errors that are a prominent source of noise occurring in long idle regions due to qubit dephasing. While this can be partially suppressed using dynamical decoupling (see Appendix), we take a further step to improve particularly bad cases by temporarily disentangling the earliest qubits and returning them to the ground state. This can be done with an as-soon-as-possible (ASAP) scheduling of the uncomputation such that as soon as the qubit has performed its task of passing on entanglement to its neighbors, it is disentangled. The relaxed qubit is well-protected for the duration of the circuit, and we only take steps to re-entangle it with the state with an as-late-as-possible (ALAP) recomputation.

By performing all of these steps, we successfully generated genuine multipartite entanglement across 120 qubits using 8 checks and 1 uncomputed qubit, with a post-selection retention rate of 0.28. To certify this entanglement, we used DFE and measured a fidelity of 0.56(3) on the $ibm\_aachen$ processor (see Fig.~\ref{fig:120}). We next describe the fidelity estimation method and present additional results on large GHZ states across several processors.

\section{Experimental Fidelity Estimation of Large-Scale GHZ States}




We use two distinct methods to estimate the fidelity of GHZ states: parity oscillations and DFE. Although both target the same quantity, i.e. the fidelity of the prepared state, they differ in their experimental implementation and sensitivity to readout errors.
In both cases, the state is said to have genuine multi-partite entanglement across all $n$ qubits if its fidelity exceeds a threshold of $0.5$~\cite{guhne2009entanglement}.

In superconducting qubit experiments, fidelity has often been estimated using the MQC protocol. This approach applies single-qubit phase shifts followed by the inverse of the preparation unitary, and infers fidelity from the ground state population. While MQC is relatively robust to readout errors~\cite{wei2020verifying}, it doubles the circuit depth and introduces non-Clifford operations that complicate the integration of error detection.

To avoid these limitations, we adopt the parity oscillations method and DFE. Parity oscillations have been widely used in trapped-ion and neutral-atom systems, where high readout fidelity supports the measurement of high-weight observables~\cite{monz201114,song2019generation}. Its application to superconducting qubits has been limited by historically lower readout fidelity and the complexity of the required measurements. Similarly, although DFE was introduced over a decade ago~\cite{flammia2011direct,da2011practical}, its implementation was previously limited by the need to perform many independent measurements of possibly high-weight Pauli observables, which was challenging due to the slow execution speed and readout fidelity of early quantum processors~\cite{bravyi2021mitigating}. Recent advances in quantum hardware speed, readout fidelities, and error mitigation~\cite{van2022model} have made it feasible to implement Parity oscillations and DFE with high precision in large-scale experiments~\cite{cao2023generation,baumer2024efficient}.

\begin{figure*}[t]
\centering
\includegraphics[width=\textwidth]{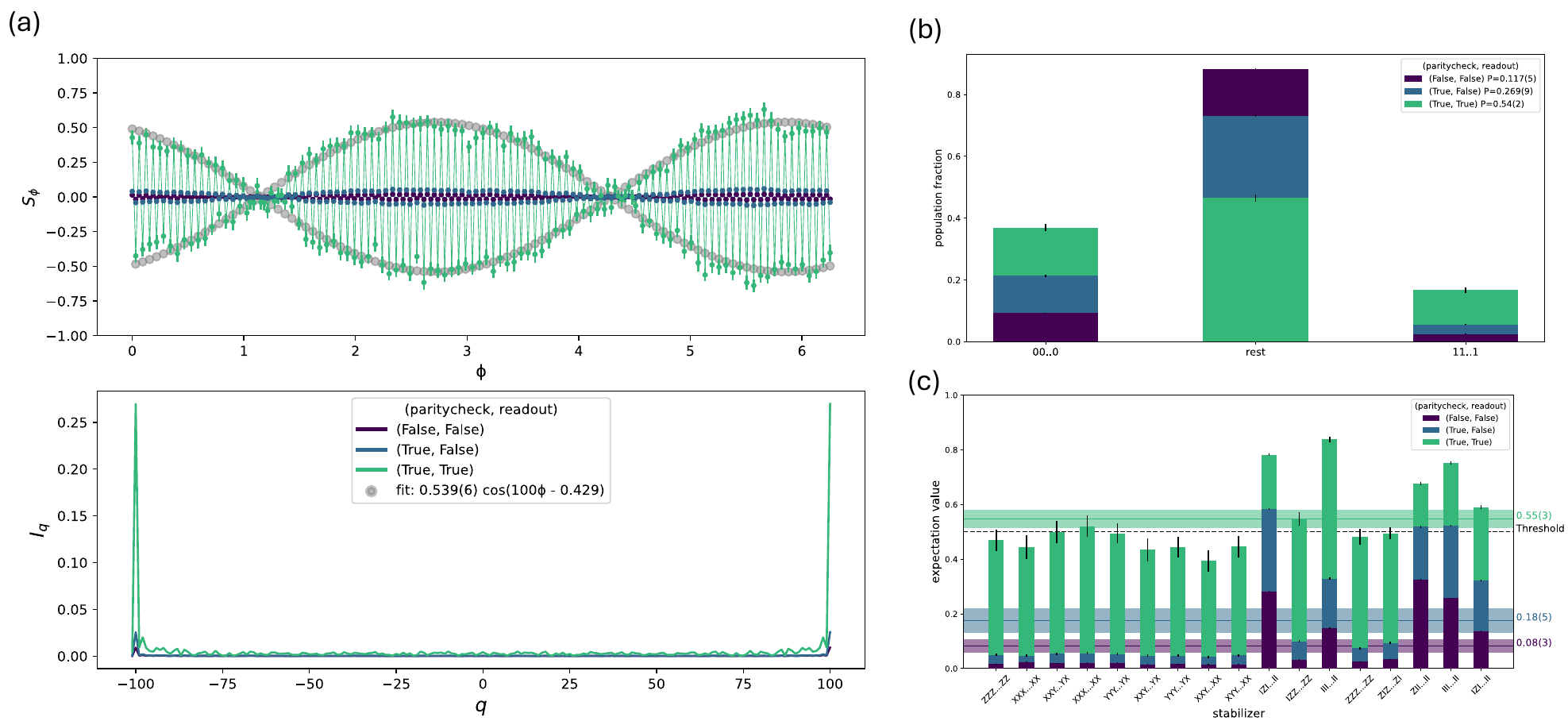} 
\caption{Demonstrating the equivalence of state certification using either (a-b) measurement of parity oscillations and population or (c) direct fidelity estimation. Identical circuits are used to prepare a 100-qubit GHZ state on $ibm\_kingston$, and three different quantities are measured. The first method gives a fidelity estimate of $F = 0.536(8)
$ (average of (a) coherence=$0.539(5)$ and (b) population=$0.54(1)$). The second method gives a direct fidelity estimate of (c) $F=0.55(3)$. The two estimates are within error bars of each other. Due to the linearly increasing number of circuits in the first method, state certification takes a longer time. Different colors in the stacked bars in panels (b) and (c) indicate how parity checks and readout mitigation improve the fidelity and its estimate, respectively.
 }\label{fig:certify}
\end{figure*}

\paragraph{Parity oscillation fidelity estimation ---}
Using the definition of fidelity and the representation of the state in the computation basis we observe that~\cite{guhne2007toolbox}
\begin{align}
    F &= \expval{\rho}{G_n} \nonumber\\
    &= \frac{1}{2}(\bra{\mathbf{0}}\rho\ket{\mathbf{0}} + \bra{\mathbf{1}}\rho\ket{\mathbf{1}} + \bra{\mathbf{0}}\rho\ket{\mathbf{1}}+\bra{\mathbf{1}}\rho\ket{\mathbf{0}}) \nonumber\\
    &= \frac{1}{2}(\expval{P}+\expval{\chi}),
\end{align}
where $\ket{\mathbf{0}}$ and $\ket{\mathbf{1}}$ have been used to denote $\ket{0}^{\otimes n}$ and $\ket{1}^{\otimes n}$, respectively. Here, $\expval{P}=\bra{\mathbf{0}}\rho\ket{\mathbf{0}} + \bra{\mathbf{1}}\rho\ket{\mathbf{1}}$ is referred to as the population and $\expval{\chi}=\bra{\mathbf{0}}\rho\ket{\mathbf{1}}+\bra{\mathbf{1}}\rho\ket{\mathbf{0}}$ as the coherence. 

The population $P$ can be estimated  directly from computational basis measurements. The coherence  $\chi$, however, needs measurements in different bases. Specifically, to estimate $\chi$, we need to measure expecation value of $Z^{\otimes n}$ after applying rotations $R_z(-\phi)$ followed by $R_y(-\pi/2)$ to all the qubits for different values of $\phi$.  Let 
\begin{equation}
    M_\phi = \prod_{j=1}^N (\cos(\phi) X_j + \sin(\phi) Y_j),
\end{equation}
denote the observable that we effectively measure after the prescribed rotations. It follows that 
\begin{equation}\label{eq:chi_m}
    \chi = \frac{1}{N}\sum_{k=1}^N (-1)^k M_{\phi_k}, 
\end{equation}
where $\phi_k = k \pi/N $ (see Appendix). 

\paragraph{Direct fidelity estimation ---}
We can also use the representation of the GHZ state in terms of its stabilizers to derive an equivalent formula for the fidelity 
\begin{align}
   F &= \expval{\rho}{G_n} \\
    &= \frac{1}{2^N}  \sum_i \Tr(\rho S_i), 
\end{align}
which is the uniform sum over  stabilizer expectation values. Therefore, while there are exponentially many terms in this sum, by uniformly sampling $S_i$ and calculating their expectation values, we can find an estimate of $F$~\cite{flammia2011direct,da2011practical}. Crucially, the sample complexity of this method does not scale with system size.

\paragraph{Equivalence between the two methods ---}
We argue that (as expected) these two methods of estimating $F$ are equivalent. Specifically, after a straightforward calculation one can observe that $P$ corresponds to the sum of all the $Z$ stabilizers of the GHZ state, while $\chi$ corresponds to all the non-$Z$ stabilizers. That is, as we show explicitly in Appendix, we have 
\begin{align}
    P&=\frac{1}{2^{N-1}} \sum_{O_i\in \mathcal{S}_d} O_i, \\
    \chi &= \frac{1}{2^{N-1}} \sum_{O_i\in \mathcal{S}\setminus \mathcal{S}_d} O_i,
\end{align}
where $\mathcal{S}_d$ is the set of all diagonal Pauli operators in $\{I,Z\}^{\otimes n}$ with an even number of $Z$. 

The parity oscillations estimate $\chi$ directly, whereas DFE samples components of them uniformly at random. Therefore, the uncertainties in estimates from the parity oscillations method originate from finite sampling of the corresponding observables, whereas the uncertainties in DFE mainly come from sampling a finite number of stabilizers.


Given the considerations above and the different assumptions in the readout mitigation strategies, we find it instructive to estimate the fidelity of a prepared GHZ state using both methods. For parity oscillations, instead of using Eq.~(\ref{eq:chi_m}) we extract $\braket{\chi}$ via $\braket{\chi}=I_N + I_{-N}$, where $I_q=\frac{1}{2(N+1)}\sum_{j=0}^{2N+1}e^{\frac{iqj\pi}{N+1}}\braket{M_{\frac{iqj\pi}{N+1}}}$ is the Fourier transform of the parity oscillation signal~\cite{bao2024creating,wei2020verifying}. Measuring coherence this way allows us to extract the phase of the GHZ state, given by $\theta=-\arctan[\text{Im}(I_N)/\text{Re}(I_N)]$. Indeed, any rotation to the GHZ state will cause $\braket{\chi}$ to decrease. However, while the fidelity with respect to the ideal GHZ is lower, the state is still multipartite entangled provided that the absolute value of the coherence $C=|I_N|+|I_{-N}|$ is above $1/2$. The parity oscillation and its Fourier transform are shown in Fig.~\ref{fig:certify} (a). In our 100-qubit experiment we find a phase offset of approximately 0.429 in our GHZ state, leading to a fidelity of $0.513(8)$ with respect to the ideal GHZ state and the maximum fidelity with respect to the rotated GHZ state is $0.536(8)$. In Fig.~\ref{fig:certify} (c), we show the results from DFE and obtain $F=0.55(3)$. Both approaches yield fidelity estimates above 0.5 and agree within one standard deviation. This agreement provides strong evidence that we have successfully prepared a GHZ state with genuine multipartite entanglement, and that the two methods behave well experimentally.

Having established confidence in our technique and fidelity estimation methods, we proceed to benchmark the preparation of 100-qubit GHZ states across several quantum processors. The results, summarized in Table~\ref{tab:devices}, report the DFE measured fidelity and post-selection rates for $ibm\_kobe$, ${ibm\_kingston}$, and ${ibm\_fez}$, where circuits were compiled for each separately. These results demonstrate the reproducibility of our approach across different processors and establish our protocol as a reasonable benchmark of large-scale quantum computers.


\begin{table}[t]
\centering
\setlength{\tabcolsep}{4pt} 
\renewcommand{\arraystretch}{1.1} 
\begin{tabular}{lcc}
\hline\hline
QPU & Fidelity & PS rate \\
\hline
$ibm\_kobe$ & 0.70(4) & 0.36 \\
$ibm\_kingston$ & 0.55(3) & 0.28 \\
$ibm\_fez$ & 0.46(4) & 0.14 \\
\hline\hline
\end{tabular}
\caption{Benchmarking 100-qubit GHZ states on multiple processors using DFE.}
\label{tab:devices}
\end{table}

\section*{Outlook}

We have demonstrated, through a combination of optimized compilation and error detection, that large resource states for quantum computation can be prepared on current generation of quantum processors, indicating that a substantial fraction of these large processors can be fully entangled.  Beside their benchmarking utility, such GHZ states can readily be used in various quantum algorithms~\cite{yoshioka2025krylov,chen2025nishimori}. While we have exploited particular structures of the GHZ state, it would be interesting to consider other stabilizer states as well. Such states may not admit low-weight stabilizers anymore, making their parity measurements more challenging. However within the bulk of the circuit one might hope to find low-weight spacetime stabilizers that could be leveraged in combination with our techniques. Post-selected creation and injection of resource states can be scalable when combined with gate teleportation techniques~\cite{delfosse2025low}.

There are other types of post-selections that one may perform to boost performance yet more. For example Table~\ref{tab:devices} reports a 100-qubit GHZ state fidelity of 0.46(4) on $ibm\_fez$, with a $0.14$ retention rate. The fidelity can be increased by incorporating a type of post-selection that targets the presence of non-Markovian errors on this device~\cite{jaytalk}. Doing this we obtain a fidelity of 0.68(3) at a retention rate of 0.04. Such methods could in principle be used to certify even larger GHZ states, but we find that they become bottlenecked by the number of available qubits in current generations of quantum processors.

\section{Code and data availability}
All code used to compile the circuits and run the experiments, as well as the data presented, is available at \hyperlink{github.com/ajavadia/big-cats}{github.com/ajavadia/big-cats}.

\bibliographystyle{apsrev4-1}
\bibliography{references}
\pagebreak
\appendix
\setcounter{secnumdepth}{2}

\section{Computing the detecting regions}
Here we give a short proof of Lemma 1, which connects the detecting region of a GHZ parity check to a simple graph concept.

\begin{proof}[Proof]

Let the spacetime locations (or wires) of the GHZ circuit $C$ be \(\mathcal{W}=\{(q, l): q\in V,\ l=1,\dots,L\}\), and let
\[
\mathcal{W}_{\mathrm{err}} := \{(q,t)\in \mathcal{W}: t \geq l_q \}
\]
be the error-eligible locations. Here $l_q$ is the first layer where the qubit $q$ is excited. Wires outside of $\mathcal{W}_{\mathrm{err}}$ are not subject to errors in this analysis because they are in the ground state.

Back–propagate the observable $Z_i Z_j$ through $C$ and record its trace on $\mathcal{W} \setminus \mathcal{W}_{\mathrm{err}}$. Because $C$ is a CNOT tree, conjugation preserves $Z$-type operators. Two $Z$ operators cancel when they first meet, which is by definition equal to $lca(i, j)$. Therefore we can define the back-propagator of the check as 
\[
\overleftarrow{B}(i,j) \;=\; \bigotimes_{k\in S_{i,j}} Z_k \;\otimes\; I_{\overline{S_{i,j}}}.
\]

Let a single–qubit Pauli error $E_{(q,t)}$ occur at spacetime location $(q,t)$.
The measurement of $Z_i Z_j$ flips iff $E_{(q,t)}$ anti-commutes with $\overleftarrow{B}(i,j)$, which is the case if $E_{(q,t)} \in \{X, Y\}$ and the corresponding edge of $(q,t)$ lies in $S_{i,j}$.
\end{proof}

\subsection{Readout error mitigation}

Our goal is to certify the fidelity of the prepared GHZ state, since a high-fidelity state as such can then be used as a resource in various protocols. However, readout errors can dominate the process of fidelity estimation. Therefore we seek to remove them from the state certification protocol, similar to how they may be removed from a randomized benchmarking protocol whose goal is to estimate a gate's fidelity.

We perform two different kinds of readout error mitigation depending on the measurement being carried out: M3~\cite{nation2021scalable} which corrects bitstrings statistically based on a learned model, and T-REX~\cite{van2022model} which corrects expectation values based on twirling and multiplying by a baseline. Part of the contribution of this paper is to show that these two give roughly similar results. 

The TREX and M3 methods are inherently targeting two complementary scenarios. Take a diagonal operator. If it has few Pauli terms (sparse), it necessarily has many unique bitstrings when measured in the Z basis. Likewise, if the distribution of bitstrings is peaked, then it will be written as a sum of many Pauli terms. We therefore use TREX in DFE and parity oscillations experiments, while M3 is used in measurement of populations.


\section{Suppressing Z errors with dynamical decoupling}

\begin{figure*}[t]\centering\begin{tabular}{cc}\includegraphics[width=0.45\textwidth]{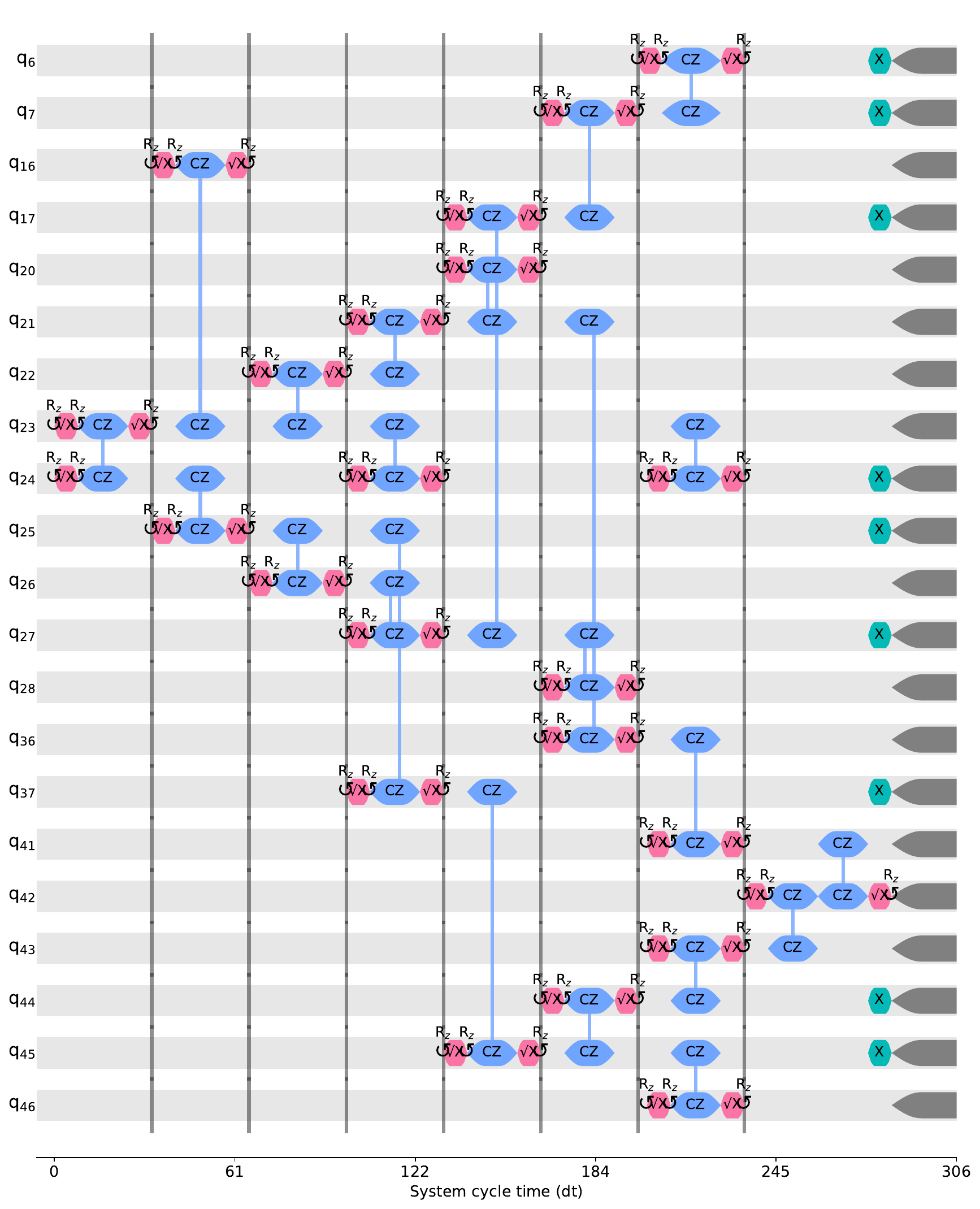} & \includegraphics[width=0.45\textwidth]{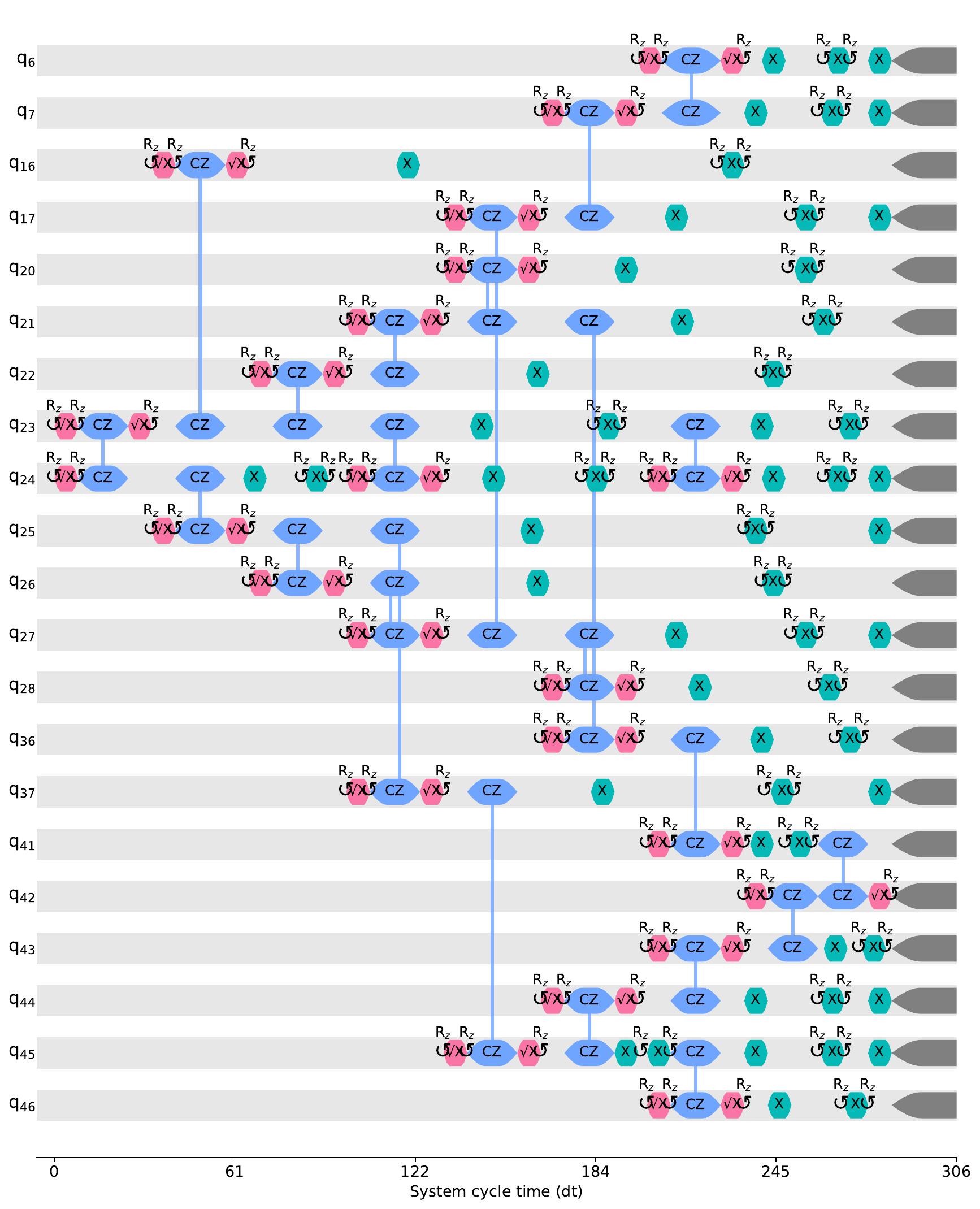} \\ \includegraphics[width=0.45\textwidth]{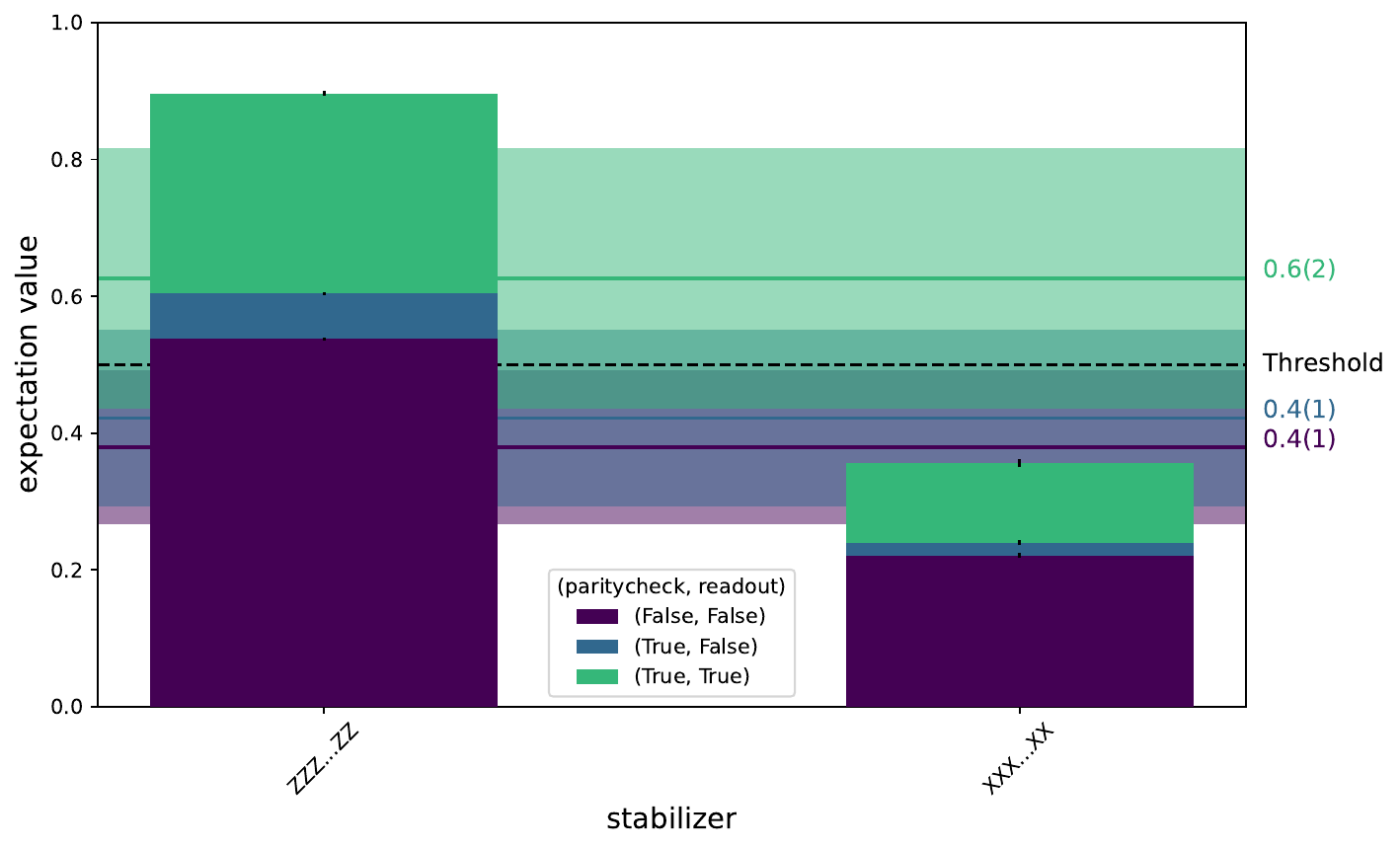} & \includegraphics[width=0.45\textwidth]{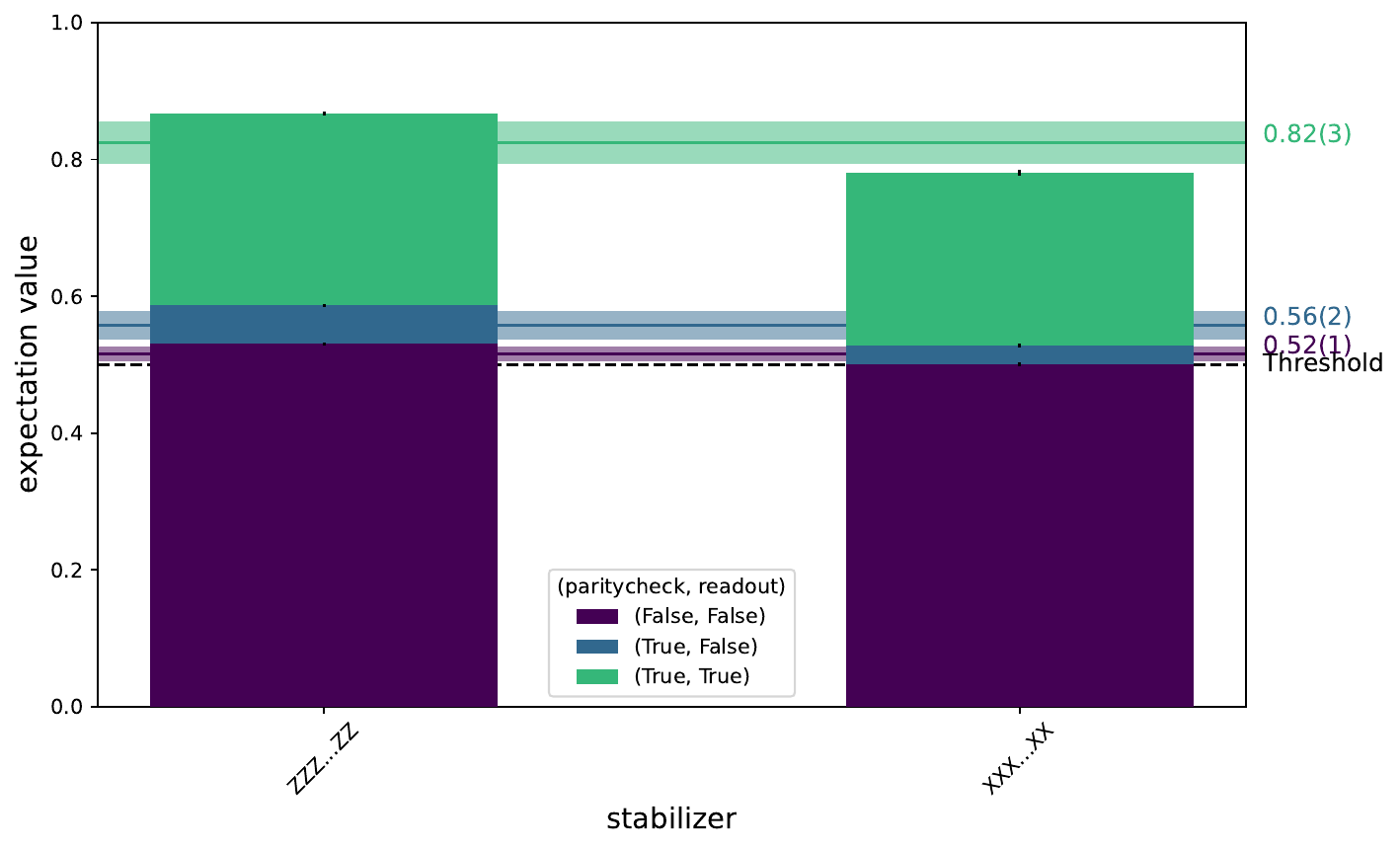} \\ \end{tabular}\caption{Effect of dynamical decoupling. The off-diagonal stabilizers are made worse by dephasing (left), unless we perform dynamical decoupling (right). The timeline view shows the exact timing of DD pulses (green) in relation to the single-qubit gates (red) and two-qubit gates (blue). The experiment prepares a 20-qubit GHZ state on $ibm\_kingston$.}\label{fig:dd}
\end{figure*}

While $ZZ$ parity checks detect $X$ or $Y$ errors, they are completely oblivious to $Z$ errors. However, $Z$ errors are commonplace, in particular due to dephasing of qubits while they are idle. Recall that the GHZ state is prepared in a tree-like fashion, and earlier qubits remain idle for a long time before the entire state is prepared, and those qubits can undergo significant dephasing to which the GHZ state is very sensitive. (Note that earlier qubits may also be measured earlier, however we intentionally measure out the qubits after the entire state is prepared, because we are interested in certifying the state and not merely the circuit.)

We use dynamical decoupling (DD) to effectively suppress dephasing errors, as shown in Figure~\ref{fig:dd}. We can confirm this by observing that lack of DD has a minimal effect on the diagonal stabilizers (it even slightly improves them due to fewer single-qubit gates), but has a dramatic worsening effect on off-diagonal stabilizers.

\section{Equivalence of fidelity estimation techniques\label{app:fidelity_derivation}}
We show how the parity oscillations scheme and DFE are connected. First, we consider estimating $P$. Note that $P = \ketbra{\mathbf{0}}{\mathbf{0}}+\ketbra{\mathbf{1}}{\mathbf{1}}$. 
We can see that 
\begin{align}
    \begin{split}
    P &= \prod_{i=1}^{N} \frac{1+Z_i}{2} + \prod_{i=1}^{N} \frac{1-Z_i}{2} \\
    &= \frac{1}{2^{N-1}} \sum_{O_i\in \mathcal{S}_d} O_i, 
    \end{split}
\end{align}
where $\mathcal{S}_d$ is the set of all Pauli operators in $\{I,Z\}^{\otimes N}$ with an even number of $Z$, that is, the subset of GHZ stabilizers that only contain $I$ and $Z$. As $P$ is the uniform sum of the expectation value of operators in $\mathcal{S}_d$, we can find an unbiased estimate of it by randomly sampling $Z$ stabilizers of GHZ.

Next, we examine $\chi= \ketbra{\mathbf{0}}{\mathbf{1}}+\ketbra{\mathbf{1}}{\mathbf{0}}$. Similar to before, we have 
\begin{align}\
\begin{split}
    \chi &= \ketbra{\mathbf{0}}{\mathbf{1}}+\ketbra{\mathbf{1}}{\mathbf{0}} \\
    &= (\ketbra{\mathbf{0}}{\mathbf{0}}+\ketbra{\mathbf{1}}{\mathbf{1}})X^{\otimes N}\\
    &=  \frac{1}{2^{N-1}} \sum_{O_i\in \mathcal{S}_d} O_i X^{\otimes N} \\ 
    & = \frac{1}{2^{N-1}} \sum_{O_i\in \mathcal{S}\setminus \mathcal{S}_d} O_i, 
\end{split}
\end{align}
that is, $\chi$ is the uniform sum of all the non-Z stabilizers of GHZ. Therefore, similar to $P$, $\chi$ can be estimated by uniformly sampling the expectation value of non-$Z$ stabilizers of GHZ. It is also clear that $\frac{1}{2} (P+\chi) = \frac{1}{2^N} \sum_{i} S_i$. 

Finally, we show explicitly that Eq.~\eqref{eq:chi_m} reduces to the sum over non-$Z$ stabilizers. We first express $M_\phi$ in a way that is easier to manipulate for this purpose
\begin{align}
\begin{split}
        M_\phi &= \prod_{j=1}^N (\cos(\phi) X_j + \sin(\phi) Y_j) \\ 
        &= \prod_{j=1}^N (\cos(\phi) X_j + i \sin(\phi) X_jZ_j) \\
        &=X^{\otimes N} \prod_{j=1}^N (\cos(\phi) I+ i \sin(\phi) Z_j)\\
        &=X^{\otimes N}  (e^{i\phi} \ketbra{0} + e^{-i\phi} \ketbra{1})^{\otimes N} \\
        &= X^{\otimes N}  \sum_{m=0}^{N} e^{2i (m-N/2)\phi}(\sum_{\pi}\ketbra{0}^{\otimes m} \ketbra{1}^{\otimes (N-m)} ),
\end{split}
\end{align}
where the inner sum is over all permutations $\pi$ of the qubits. Next, let $Q_m = \sum_{\pi}\ketbra{0}^{\otimes m} \ketbra{1}^{\otimes (N-m)} $, and plug the above expression in Eq.~\eqref{eq:chi_m}. We find
\begin{equation}
    \begin{split}
    \chi &= \frac{1}{N}\sum_{k=1}^N (-1)^k M_{\phi_k} \\
    &=\frac{1}{N} \sum_{k=1}^N (-1)^k (X^{\otimes N}  \sum_{m=0}^{N} e^{2i (m-N/2)\phi_k} Q_m)\\
    &= \frac{1}{N} X^{\otimes N}  \sum_{k=1}^N\sum_{m=0}^N  (-1)^k e^{2i (m-N/2)\phi_k} Q_m \\
    &= \frac{1}{N} X^{\otimes N}  \sum_{k=1}^N\sum_{m=0}^N  (-1)^k e^{2i (m-N/2)k \pi/N} Q_m \\
    &= \frac{1}{N} X^{\otimes N}  \sum_{k=1}^N\sum_{m=0}^N  e^{-i\pi k} e^{\frac{2i\pi }{N} (m-N/2)k } Q_m \\
    &= \frac{1}{N}  X^{\otimes N}  \sum_{k=1}^N\sum_{m=0}^N  e^{-\frac{2i\pi}{N} N k/2} e^{\frac{2i\pi }{N} (m-N/2)k } Q_m\\
    &= X^{\otimes N}  \sum_{m=0}^N  \delta_{N/2,(m-N/2)} Q_m\\
    &=  X^{\otimes N}(Q_0 + Q_N)\\
    &= X^{\otimes N}(\ketbra{\mathbf{0}} + \ketbra{\mathbf{1}})
    \end{split}
\end{equation}
where we used the orthonormality of the Fourier basis. Therefore, we again see that $\chi$ is the uniform sum over all the non-$Z$ stabilizers.

\section{Calculating error bars}
The uncertainties in fidelity estimates come from statistical errors both in the sampled stabilizers and in the samples taken to estimate each stabilizer expectation value. Therefore, to estimate the uncertainty we have to consider both sources of errors.

Let $\mathcal{S}$ denote the stabilizer expectation value and $\mathcal{X}$ denote which stabilizer we measure. We have
\[
\mathbb{E}[\mathcal{S} | \mathcal{X} = \hat{s}_i].
\]

Let $s_i, \sigma_i^2$ be the mean and variance for $\hat{s}_i$.

Define
\[
F = \mathbb{E}[\mathcal{S}].
\]

Then, from the law of total variance we have
\[
\mathrm{Var}(F) = \mathbb{E}[\mathrm{Var}(\mathcal{S} | \mathcal{X})] + \mathrm{Var}(\mathbb{E}[\mathcal{S} | \mathcal{X}]).
\]

We also have
\[
\mathbb{E}[\mathrm{Var}(\mathcal{S} | \mathcal{X})] = \mathbb{E}[\sigma_i^2],
\]
\[
\mathrm{Var}(\mathbb{E}[\mathcal{S} | \mathcal{X}]) = \mathrm{Var}(s_i).
\]

Therefore, we can find the variance of $F$ as 
\[
\mathrm{Var}(F) = \mathbb{E}[\sigma_i^2] + \mathrm{Var}(s_i).
\]

Now suppose we estimate its mean, $\bar{F}$, from $m$ samples. That is 
\[
\bar{F} = \frac{1}{m} \sum_i \xi_i.
\]

Then the variance of the mean of $F$ with $m$ samples  is given by
\[
\mathrm{Var}(\bar{F}) = \frac{1}{m^2} \left( \sum_i \mathbb{E}[\sigma_i^2] + \sum_i \mathrm{Var}(s_i) \right),
\]

which simplifies to
\[
\mathrm{Var}(\bar{F}) = \frac{1}{m} \left( \mathbb{E}[\sigma_i^2] + \mathrm{Var}(s_i) \right).
\]

\end{document}